\documentstyle{laa}
\input epsf.sty
\input psfig.sty
\begin{document}
\thesaurus{02.13.2;08.16.6;09.10.1}
\title{Magnetocentrifugal acceleration of plasma in a 
nonaxisymmetric magnetosphere}
\author{S.V.Bogovalov}
\institute{Moscow Engineering Physics Institute, Kashirskoje Shosse 31,Moscow, 
115409, Russia. e-mail: bogoval@axpk40.mephi.ru}
\date{Received January 1, 2000; accepted November 3, 2000}

\maketitle
\begin{abstract}
Violation of the axial symmetry of a magnetic field
essentially modifies the physics of the plasma outflow
in the magnetosphere of rotating objects. In comparison to
the axisymmetric outflow, two new effects appear: more efficient
magnetocentrifugal acceleration of the plasma along particular field
lines and generation of  MHD waves.
Here, we use an ideal MHD approximation to study the dynamics of a
cold wind in the nonaxisymmetric magnetosphere.
We obtain a self-consistent analytical solution of the problem of 
cold plasma
outflow from a slowly rotating star with a slightly nonaxisymmetric
magnetic field using perturbation theory.
In the axisymmetric (monopole-like) magnetic field, the first term in the
expansion of the terminating energy of the plasma in powers of $\Omega$
is proportional to $\Omega^4$, where $\Omega$ is the angular
velocity of the central source. 
Violation of the axial symmetry of the magnetic field crucially changes this
dependence. The first correction to the energy of the plasma
becomes proportional to $\Omega$.
Efficient magnetocentrifugal acceleration occurs along the
field lines curved initially in the direction of the rotation.
I argue  that  all the necessary conditions for
the efficient magnetocentrifugal acceleration of the plasma exist in
 the radio pulsar magnetosphere. We calculated  the
first correction of the rotational losses due to the generation of the
MHD waves  and
analysed  the plasma acceleration by these waves.

\end{abstract}

\section{Introduction}

The outflow of plasma from rotating magnetised objects is a widespread
phenomenon
in the Universe. It is observed in a wide range of astrophysical objects 
of different natures and scales. Some of these objects eject relativistic 
plasma. The Lorentz factor of plasma ejected by microquasars and AGN
is of the order 10 (Mirabel \& Rodrigues \cite{mirabel}, Begelman et al.
\cite{bbr}, Pelletier et al. \cite{pelletier}), although there is evidence 
that in Blazars the plasma is accelerated to the Lorentz factor
$\sim 10^6$ (Aharonian et al. \cite{ah}). Radio pulsars also 
accelerate the plasma to the Lorentz factor $\sim 10^6$ 
(Kennel \& Coroniti \cite{kennel}, Coroniti \cite{coroniti}, Arons 
\cite{arons96}). The problem of the acceleration of the 
plasma to such high
energies  remains one of the most important  unsolved problems in  
modern relativistic astrophysics.

The acceleration of plasma can occur in principle
due to different dissipative as well as nondissipative processes.
Hypothesis have been proposed  about the singular behaviour of the
relativistic plasma outflow
at the ``light surface'' (\cite{beskin83}) or at the light cylinder
(Mestel \& Shibata \cite{ms}) of radio pulsars. These dissipative structures
could provide  the  plasma
acceleration. However, recent direct  numerical
calculations of the relativistic outflows in MHD
approximation (Bogovalov \cite{bog97}) and force-free approximation
(Contpoulos et al. \cite{janis}) have
demonstrated that the relativistic plasma flow is actually
regular everywhere.

Another mechanism of  acceleration due to dissipative processes
was proposed by Michel (1982) and
Coroniti (1990) in application to radio pulsars.
Michel was the first who noticed that the number of charge carriers in the
wind from radio pulsars  is not enough to sustain the necessary return
current in the wind with a striped magnetic field formed at the outflow from
the oblique rotator.  Coroniti (1990) considered
the reconnection process in this magnetic field and came to the conclusion
that it can provide the necessary acceleration of
the winds from radio pulsars. Up to now this process was considered as the
only possible mechanism explaining pulsar wind acceleration. However, recently
Luybarskii \& Kirk (2000) have revisited  the process of wind acceleration
due to  magnetic field reconnection. They  concluded that well before
the winds from pulsars such as the Crab pulsar reach high Lorentz factors
they will be terminated by the interstellar medium.

A lot of efforts has been spent to obtain the
plasma acceleration in frameworks of ideal MHD.
There is a general opinion that in all astrophysical objects ejecting
plasma, the magnetic field plays an essential role in the dynamics
of the plasma. In particular, the magnetic field of a rotating object
itself
can provide the acceleration of the plasma due to a magnetocentrifugal 
(below MC)
mechanism (Blandford \& Payne \cite{bp}). The ideal MHD outflows of 
plasma in different models  and
different approximations were investigated in a series of works 
(Michel \cite{michel69}, Sulkanen \& Lovelace \cite{lovlace},
Ferreira \cite{ferreira}, Vlahakis 
\& Tsinganos \cite{tsin},  Bogovalov \& Tsinganos \cite{bogtsin},
Krasnopolsky  et al. \cite{kb}, Ustyugova et al. \cite{ustyug}).
It follows from these studies that nonrelativistic plasma can be 
accelerated rather efficiently due to this mechanism. As far as 
relativistic plasma is concerned,  the efficiency of the 
acceleration of this plasma  appears insufficient for astrophysical 
applications.
There were hopes that divergence of the magnetic flux tubes somewhere 
beyond the  light cylinder will result in more efficient acceleration of the 
plasma due to the pressure gradient of the toroidal magnetic field 
(Begelman \& Li \cite{bl}, Takahashi \& Shibata \cite{takshib}).  
The only known mechanism for this 
magnetic flux tube divergence  is  magnetic self-collimation 
of the magnetised winds (Heyvaerts \& Norman \cite{heyverts},
Chiueh et al. \cite{chuk}). This mechanism can, in principle,
provide divergence 
of the field lines at the equatorial plane such that the poloidal field 
decreases with distance faster than $r^{-2}$. This divergence can take
place in the limiting range of distances from the source, since
finally the wind  expands radially near the equator at
$r\rightarrow \infty$  (Heyvaerts \& Norman
\cite{heyverts}, Chiueh et al. \cite{chuk}).
   
Numerical self-consistent solution of the problem of  
steady-state  plasma outflow in a model with the an initially monopole-like 
magnetic field shows that collimation really increases the efficiency 
of nonrelativistic plasma acceleration at the equator
(Bogovalov \& Tsinganos \cite{bogtsin}). 
However, collimation of the  relativistic plasma outflow  appears negligibly 
small at large
Lorentz factors  (Bogovalov \cite{bog97}, Bogovalov \cite{bog00}).
Therefore, the relativistic plasma is 
practically not accelerated by the pressure gradient of the toroidal
magnetic field. This conclusion is valid not only for the
monopole-like model.

 The acceleration of the wind due to the pressure gradient of the 
toroidal magnetic field occurs at the large distances compared to the 
dimension of the central source (Begelman \& Li \cite{bl},
Takahashi \& Shibata \cite{takshib}). 
The stationary wind from any source is monopole-like at these distances. 
Therefore, the poloidal 
magnetic field  also becomes monopole-like, 
since the magnetic field is frozen 
into the plasma. Thus, the solutions obtained in the monopole-like 
model actually describe general properties of any axisymmetric wind at 
large distances from the source.

The application of the  monopole-like model is not limited by large 
distances. This model  also gives  upper  limits  on  the  efficiency
of  the MC acceleration 
of the plasma in any other axisymmetric  magnetic field 
(dipolar for example).
It is qualitatively clear that the slower the magnetic field
decreases with  distance, the more efficiently it accelerates
the plasma due to a MC mechanism. The monopole-like magnetic field  falls
down  as  $r^{-2}$,
while the dipolar magnetic field falls down faster than $r^{-2}$. 
Correspondingly, the plasma is expected to be more efficiently accelerated 
in the monopole-like magnetic field in compare with the acceleration in the
dipolar one.
This is why the results obtained in the monopole-like model rule out
efficient MC acceleration of the plasma in
any other more realistic axisymmetric model.
Calculations  performed by
Contopoulos et al. (\cite{janis}) confirm this conclusion.

Thus, up to now all the attempts to propose an effective
mechanism of relativistic plasma acceleration have not been successful.
That is why the search for new possible mechanisms to explain
relativistic plasma acceleration remains one of
the most important problems of  relativistic astrophysics.

The disappointing conclusion regarding the low efficiency of the MC
acceleration  is valid only for   axisymmetric
magnetic fields, as previous studies considered  this process
 only within the frameworks of  axisymmetric models.
However,  the magnetic field of  real astrophysical objects is not
axisymmetric. Radio pulsars give us the  brightest example of this 
relation. Therefore, it is important to know how  the
non axisymmetry of the magnetic field affects the process of  MC
acceleration. In this paper we try to answer this question.

To clarify the role of the nonaxisymmetry of the magnetic field for the 
MC  acceleration  of    plasma,  a  model  with  the
initially monopole-like magnetic field is used in this paper.
This model is often used to investigate the processes of 
plasma collimation and acceleration  in the rotating
magnetosphere (Michel \cite{michel69}, Mestel \& Selley \cite{mestel},
Sakurai  \cite{sakurai},   Bogovalov   \cite{bog92}, Beskin et al. 
\cite{vasia98},
Bogovalov  \&  Tsinganos  \cite{bogtsin},  
Tsinganos   \& Bogovalov \cite{tsbog}). 
The word ``initially" implies that
the magnetic field of the non rotating star looks like the magnetic field
of the ``magnetic monopole".
This model is very convenient since there are no closed field lines in the
flow. It remarkably simplifies the analysis of  the
plasma acceleration.

The main goal of
this paper is to  answer  the question: how does the nonaxisymmetry of the
magnetic field of a star affect the process of  plasma acceleration?
An attempt to solve the problem
of the structure of the magnetosphere of an oblique rotator with a dipole
magnetic field in mass-less approximation 
has been done by Beskin et al. (\cite{beskin83}). 
The first step in the solution of this
problem in the MHD approximation has been done in the work of Bogovalov
(\cite{bog99}). In this work, the
problem of plasma flow in the magnetosphere of the oblique rotator with
an initially split-monopole magnetic field was solved. 
The modulus of the magnetic field was axisymmetric and only the direction
of the magnetic field varied with time and the azimuthal angle.
It was found that
the acceleration remains non efficient in this flow as well. In the present
paper,
we are interested in the affect of the nonaxisymmetry of the  modulus of the
magnetic field on the process of  plasma acceleration.

Plasma flow is described by a system of non linear equations in partial
derivatives. Usually  these equations can be integrated only
numerically.
There is one exception from this rule.  Sometimes the problem can be
solved self-consistently and
analytically if we are interested in small corrections to the known solution.
These corrections can arise if the known flow is perturbed slightly by
additional  small forces. In particular,  slow rotation can be
considered as a small perturbation of some known flow from the non rotating
star. This
approach was firstly successfully applied for the numerical
self-consistent solution of
the flow of the solar wind by Nurney and Suess (\cite{ns}). Later
Bogovalov (\cite{bog92}) has used this approach for the fully analytical,
self-consistent solution
of the problem of the cold plasma outflow, relativistic as well as 
nonrelativistic,
from a slowly rotating star. In recent years, this method has also been
used to solve  a range of problems by Beskin
(see Beskin \& Okamoto \cite{bsknokmto}) and references therein).
The idea of the present work is the following: if the nonaxisymmmetry of the
magnetic field modifies the process of 
MC acceleration of the plasma, this modification can be
seen at the level of the slow rotation of the object.
 Therefore, to answer the
 question: does  the nonaxisymmetry of the
magnetic field modify the process of  acceleration of the plasma,
we investigate the acceleration of  plasma in the nonaxisymmetric
magnetic field at slow rotation.

This paper is organised as follows. In Sect. 2  we present ideal
MHD equations defining the dynamics of cold relativistic plasma.
A self-consistent analytical solution of the problem is given in
Sec. 3. The solution in the wave zone is given in Sect. 4.
The possible astrophysical implications of the results are  discussed
in Sec. 5.

\section{Basic equations and relationships}

The system of   time dependent equations defining the temporal evolution
of the relativistic plasma outflow  in an ideal MHD approximation is as follows
(Akhiezer et al. \cite{akhiezer}):
\begin{equation}
mn({\partial \gamma {\bf v}\over \partial t}+({\bf v\nabla})\gamma {\bf v})
=q\cdot {\bf E} +{1\over c} ~{\bf j}\times {\bf H},
\end{equation}
 \begin{equation}
{\partial {\bf H}\over c\partial t} = -curl  ~{\bf E},
\end{equation}
 \begin{equation}
curl ~{\bf H} ={4\pi \over c}{\bf j} +{\partial {\bf E}\over c\partial t},
\end{equation}
\begin{equation}div~{\bf H} =0,\end{equation}
\begin{equation}div~{\bf E} =4\pi q,\end{equation}
\begin{equation}{\partial n\over \partial t}+ div ({\bf v}n) =0.
\label{6}
\end{equation}
\begin{equation}{\bf E}+{1\over c} {\bf v}\times {\bf H}=0.
\end{equation}
Gravitation is neglected in these equations.
It is convenient to consider the plasma flow in a spherical coordinate
system.
In this system, the equations of motion are
\begin{eqnarray}
\lefteqn{mn({\partial \gamma v_r\over \partial t}+({\bf v\nabla})\gamma v_r-
{\gamma(v_{\theta}^2+v_{\varphi}^2)\over r})=}\nonumber\\
&&q\cdot E_r +{1\over 4\pi}\Bigl\{({1\over r\sin{\theta}}{\partial 
H_r\over\partial\varphi}
-{1\over r}{\partial(rH_{\varphi})\over \partial r})H_{\varphi}-\nonumber\\
&& -({\partial(rH_{\theta})\over r \partial r}-{\partial H_r\over r \partial 
\theta})H_{\theta}+
{1\over c}(H_{\theta}{\partial E_{\varphi}\over \partial t}-H_{\varphi}{\partial 
E_{\theta}\over \partial t})\Bigr \},
\label{1}
\end{eqnarray}
\begin{eqnarray}
\lefteqn{mn({\partial \gamma v_{\theta}\over \partial t}+({\bf v\nabla})\gamma 
v_{\theta}+
{\gamma (v_{r}  v_{\theta}-v_{\varphi}^2 \cot\theta ) \over r})=}\nonumber\\
&&q\cdot E_{\theta} +{1\over 4\pi}\Bigl\{({\partial (rH_{\theta})\over r\partial 
r}
-{\partial H_{r}\over r\partial \theta})H_{r}-\nonumber\\
&& -{H_{\varphi}\over r\sin\theta}({\partial \sin\theta H_{\varphi}\over  
\partial \theta}-
{\partial H_{\theta}\over \partial \varphi})+\nonumber\\
&&+{1\over c}(H_{\varphi}{\partial E_{r}\over \partial t}-H_{r}{\partial 
E_{\varphi}\over
\partial t})\Bigr\},
\label{2}
\end{eqnarray}
\begin{eqnarray}
\lefteqn{mn({\partial \gamma v_{\varphi}\over \partial t}+({\bf v\nabla})\gamma 
v_{\varphi}+
{\gamma (v_r v_{\varphi}+v_{\theta}v_{\varphi}\cot\theta)\over r})=}\nonumber\\
&&q\cdot E_{\varphi}
+{1\over 4\pi}\Bigl\{{1\over r\sin\theta}({\partial \sin\theta H_{\varphi}\over
\partial \theta}-{\partial H_{\theta}\over \partial \varphi})H_{\theta}-
\nonumber\\
&& -({1\over r\sin\theta}{\partial H_{r}\over  \partial \varphi}-
{\partial rH_{\varphi}\over r\partial r})H_{r}+\nonumber\\
&&+{1\over c}(H_{r}{\partial E_{\theta}\over \partial t}-H_{\theta}{\partial 
E_{r}\over
\partial t})\Bigr\},
\label{3}
\end{eqnarray}

and  the induction equations:
\begin{equation}
{\partial H_r\over c\partial t}=-{1\over r\sin\theta}({\partial (\sin\theta 
E_{\varphi})\over r \partial \theta}-{\partial E_{\theta}\over 
\partial\varphi}),
\label{i1}
\end{equation}
\begin{equation}
{\partial H_{\theta}\over c\partial t}=-({1\over r\sin\theta}{\partial 
E_{r}\over r \partial \varphi}-{\partial (rE_{\varphi})\over r\partial r}),
\label{i2}
\end{equation}
\begin{equation}
{\partial H_{\varphi}\over c\partial t}=-({\partial (rE_{\theta})\over r 
\partial r}
-{\partial E_{r}\over  r\partial\theta}).
\label{i3}
\end{equation}
The laws of conservation of the magnetic field  and  the matter flux,
together with the Coulomb law, take the following form:
\begin{equation}
{1\over r^2}{\partial (r^2H_r)\over \partial r}+{1\over 
r\sin\theta}{\partial\sin\theta H_{\theta}\over \partial\theta}+
{1\over r\sin\theta}{\partial H_{\varphi}\over \partial\varphi}=0.
\label{m1}
\end{equation}
 \begin{equation}
{\partial n\over \partial t}+{1\over r^2}{\partial (r^2nv_r)\over \partial 
r}+{1\over r\sin\theta}{\partial\sin\theta nv_{\theta}\over \partial \theta}+
{1\over r\sin\theta}{\partial nv_{\varphi}\over \partial\varphi}=0.
\label{n1}
\end{equation}
\begin{equation}
{1\over r^2}{\partial (r^2E_r)\over \partial r}+{1\over 
r\sin\theta}{\partial\sin\theta E_{\theta}\over \partial \theta}+
{1\over r\sin\theta}{\partial E_{\varphi}\over \partial\varphi}=4\pi q.
\label{c1}
\end{equation}

The ideal MHD conditions are
\begin{equation}
E_r+{1\over c} (v_{\theta}H_{\varphi}-v_{\varphi}H_{\theta})=0,
\label{f1}
\end{equation}
\begin{equation}
E_{\theta}+{1\over c} (v_{\varphi}H_{r}-v_{r}H_{\varphi})=0,
\label{f2}
\end{equation}
\begin{equation}
E_{\varphi}+{1\over c} (v_{r}H_{\theta}-v_{\theta}H_{r})=0.
\label{f3}
\end{equation}
In this system,  $q$ is the induced space electric charge density,
$\theta$ is the polar angle
and $\varphi$ is the azimuthal angle.

The electric field in the axisymmetrically rotating steady state
magnetosphere is connected with the poloidal magnetic field as follows:
${\bf E} = r\sin\theta\Omega/c {\bf B_p\times e_{\varphi}}$ and
$E_{\varphi}=0$ (Weber \& Davis \cite{weber}). For
the nonaxisymmetric
magnetic field the same relationship is also valid, provided that the
flow is in the steady state  and all the variables vary periodically 
with time  (Beskin et al. \cite{beskin83}).  In the spherical
coordinates, the relationships  take the form
\begin{equation}
E_{\theta}=-{r\sin\theta\Omega\over c} H_r,
\label{e1}
~~E_{r}={r\sin\theta\Omega\over c} H_{\theta},~~E_{\varphi}=0.
\end{equation}
It is easy to show that these
relationships indeed satisfy all the equations of the electric field in a
rotating magnetosphere.
At  stationary rotation, all the variables depend  on $\varphi$ and $t$
in the spherical system of coordinates as the difference
$\xi=\varphi-\Omega t$.
Then,  the induction equation (\ref{i1}) is reduced to the
equation
\begin{equation}
{\partial\over \partial\xi}({E_{\theta}\over r\sin\theta}+ {\Omega H_r\over 
c})=0
\end{equation}
which  is fulfilled due to  Eq. (\ref{e1}). The same is valid for the
induction equation (\ref{i2}) which is reduced to the equation
\begin{equation}
{\partial\over \partial\xi}({E_{r}\over r\sin\theta}- {\Omega H_{\theta}\over 
c})=0.
\end{equation}
The third   induction equation (\ref{i3})
after the substitution of  Eqs. ({\ref{e1}) is reduced
to the magnetic flux conservation condition (\ref{m1}).
Eq. (\ref{e1}) also satisfies the boundary
conditions on the surface of the star where the tangential components of the
electric field should be continuous. It follows from the ideal MHD conditions
(\ref{f1}-\ref{f3}) that below the surface of the star the tangential
components of the electric field are $E_{\varphi}=0$ and
$E_{\theta}=-\Omega r\sin\theta/c H_r$. They exactly equal the tangential
components of the electric field directly above the star surface, given by
(\ref{e1}).

\section{Acceleration of the plasma at slow rotation}

\subsection{Acceleration in the axisymmetric magnetic field}

The solution of the problem of  steady state plasma outflow can be
expanded in powers of $\Omega$. The poloidal magnetic field, the plasma
density, the poloidal velocity and the Lorentz factor of the plasma
are invariant in relation to the rotation direction at the axisymmetric
outflow. Therefore they  are expanded in even powers of $\Omega$.
Correspondingly, the toroidal magnetic field and the toroidal velocity are
expanded in odd powers of $\Omega$, since these variables change sign at the
change of the direction of rotation. Therefore, the expansion of the
solution in the powers of $\Omega$ takes the  form
\begin{equation}
{\bf H}_p={\bf H}_{p,0}+\Omega^2{\bf H}_{p,2}+ ....,
\end{equation}
\begin{equation}
H_{\varphi}=\Omega H_{\varphi,1}+\Omega^3H_{\varphi,3}+ ....,
\end{equation}
\begin{equation}
{\bf v}_p ={\bf v}_{p,0}+\Omega^2{\bf v}_{p,2}+ ....,
\end{equation}
\begin{equation}
v_{\varphi}=\Omega v_{\varphi,1}+\Omega^3 v_{\varphi,3}+ ....,
\end{equation}
\begin{equation}
n= n_{0}+\Omega^2 n_2 + ....,
\end{equation}
\begin{equation}
\gamma=\gamma_0+\Omega^2 \gamma_2+\Omega^4 \gamma_4 + ......,
\end{equation}
where lower index ``0" denotes the variables for the nonrotating star.
The first corrections to the unperturbed cold plasma outflow from
the nonrotating star arising at  slow rotations of the star with
a monopole-like magnetic field were
calculated analytically by Bogovalov (\cite{bog92}). The  first
order corrections to the  toroidal
magnetic field and the toroidal velocity are as follows
\begin{equation}
 H_{\varphi,1}=-H_{p,0} {r \sin\theta\over v_0}, ~~~~v_{\varphi,1}=0,
\label{bogsol}
\end{equation}
where
\begin{equation}
H_{p,0}= H_*({R_*\over r})^2
\label{mon}
\end{equation}
is the monopole-like magnetic field of the star and the lower index ``*''
denotes
the values on the surface of the star.
It follows from (\ref{bogsol}) that the first correction to the energy of
the plasma accelerated in the monopole-like magnetic field 
is proportional to $\Omega^4$. Indeed, the
equation for the  Lorentz factor of the plasma has the form 
(Landau \& Lifshitz \cite{landau})
\begin{equation}
mnc^2({\partial \gamma\over\partial t} +({\bf v\nabla})\gamma)=
{\bf jE}.
\label{gamma}
\end{equation}
The second order correction to the Lorentz factor  is defined as follows
\begin{equation}
mn_0c^2v_0\Omega^2{\partial\over r}\gamma_2=
{r\Omega\sin\theta\over 4\pi}H_{p,0}
{\partial\over r\partial r}(r H_{\varphi,1}).
\end{equation}
But according to (\ref{bogsol}) and (\ref{mon})
the right hand part of this equation is
equal to zero. Thus, $\gamma_2=0$. The  fact that the first
correction to the Lorentz factor at the axisymmetrical outflow begins
with the term proportional to $\Omega^4$ reflects the very
low efficiency of the MC  acceleration in the monopole-like magnetic field.

\subsection{Acceleration in the nonaxisymmetric magnetic field}

The Lorentz factor of
the plasma accelerated in the nonaxisymmetric magnetic field
is already not invariant in relation to the direction of the rotation.
In this case it can (but not must)
depend on the sign of $\Omega$. The expansion of the Lorentz factor in
$\Omega$ takes the general form
\begin{equation}
\gamma=\gamma_0 +\Omega\gamma_1+\Omega^2\gamma_2+\Omega^3\gamma_3+...
\end{equation}
In the axisymmetric case, $\gamma_1=\gamma_3=0$ (generally speaking,
the term $\gamma_2=0$ applies only for the monopole-like 
magnetic field).
In the nonaxisymmetric case these terms
can (but not must) differ from zero. If for example $\gamma_1 \ne 0$,
it would imply that the acceleration of the plasma in the nonaxisymmetric
magnetic
field is essentially more efficient in comparison to the acceleration in the
axisymmetric magnetic field, at least for the case of slow rotation. To make 
sure
that indeed $\gamma_1 \ne 0$, it is sufficient to do this for the
magnetic field having small nonaxisymmetry. In this case, the problem can
be solved analytically.

\subsection{Outflow at $\Omega =0$ in the slightly nonaxisymmetric magnetic 
field}

Let us initially consider the cold plasma outflow from 
the nonrotating star with a slightly nonaxisymmetric magnetic field.
It follows from (\ref{gamma}) that the energy of the cold plasma is
constant since the electric
field ${\bf E}=0$ everywhere. For the slightly nonaxisymmetric magnetic field
the solution also can be expanded on a small parameter, $\varepsilon$,
characterising the nonaxisymmetry of the initial split-monopole
magnetic field.
This expansion has the form
\begin{equation}
{\bf H}_p={\bf H}_{p,0}+\varepsilon{\bf h}_{p,1}+ ....,
\end{equation}
\begin{equation}
H_{\varphi}=\varepsilon h_{\varphi,1}+\varepsilon^2h_{\varphi,2}+...
\end{equation}
\begin{equation}
{\bf v}_p ={\bf v}_{p,0}+\varepsilon{\bf V}_{p,1}+\varepsilon^2{\bf V}_{p,2}
+ ....,
\end{equation}
\begin{equation}
v_{\varphi}=\varepsilon V_{\varphi,1}+\varepsilon^2 V_{\varphi,2}+ ....,
\end{equation}
\begin{equation}
n= n_{0}+\varepsilon \tilde n_1 + ....,
\end{equation}
but
\begin{equation}
\gamma=\gamma_0,
\end{equation}
since the energy of the cold plasma is constant.
It is convenient to consider the perturbation of the magnetosphere
due to a non uniform distribution 
 of the magnetic field on the stellar surface. We
assume for simplicity that the magnetic field
has its own axis of symmetry ${\bf m}$ not coinciding with the axis of 
rotation, as  is shown in Fig. \ref{fig1}.
It is convenient  firstly to solve the problem in
the coordinate system where the magnetic field is axially symmetric
and then to transform the solution to the laboratory coordinate system.
The equation for the first corrections in the system where the magnetic
field is axisymmetric follows from Eq. (\ref{2}) and takes the form
\begin{equation}
mn_0\gamma_0v_0{\partial rV_{\chi,1}\over r\partial r}=
{H_0\over 4\pi}({\partial rh_{\chi,1}\over r\partial r}-
{\partial h_{r,1}\over r\partial\chi}),
\end{equation}
where $\chi$ is the polar angle in the system of the symmetry of the flow.
It is convenient to introduce the flux function $\psi$. The components of
the poloidal magnetic field can be expressed through the flux function as
follows
\begin{equation}
H_{\chi}=-{1\over r\sin\chi}{\partial \psi\over\partial r};~~~
H_r={1\over r\sin\chi}{\partial \psi\over r \partial \chi}.
\end{equation}
The expansion of the  function $\psi$ in powers of $\varepsilon$ is
\begin{equation}
\psi=H_*R^2_*(1-\cos\chi +\varepsilon f({\bf r})+....).
\end{equation}
Taking into account that  due to the ideal MHD 
condition ${\bf v}= v {\bf H}/H$, it is easy to obtain the
equation for $f$ (Bogovalov \cite{bog92})
\begin{equation}
f_{rr}+{2\over r}f_r=({R_a\over r})^2(f_{rr}+
{(1-\eta^2)\over r^2}{\partial^2 f\over\partial^2\eta}),
\end{equation}
where $\eta =\cos\chi$, $R_a=\sqrt{H_*^2R_*^2/4\pi m n_* \gamma_0v_0^2}$ is
the Alfven (or fast magnetosonic) radius of the unperturbed flow.
The perturbation $f$ can be presented as
\begin{equation}
f=\sum_m f_m (r) Q_m(\eta),
\end{equation}
where $Q_m$ are the eigenfunctions of the differential equation
$(1-\eta^2)\partial^2 f/\partial^2\eta Q_m =-m(m+1)Q_m$. They have
the form $Q_m(\eta)= (1-\eta^2)\partial P_m(\eta)/\partial\eta$, where
$P_m $ are the Legendre polynomials of the order m.

The equation for $f_m$ takes the form
\begin{equation}
(x^2-1){d^2f_m\over d^2 x}+2x {d f_m\over d x}+{m(m+1)\over x^2}f_m=0,
\label{eqf}
\end{equation}
where $x=r/R_a$.
The general solution can be presented as a sum
$
f_m=a_m Z_m(x)+b_m Y_m(x)
$
of two independent solutions. One of them, $Z_m$, is regular at the point
$x=1$ and another, $Y_m$, is singular at this point. The condition of 
regularity of the solution at this point gives  $b_m=0$.
The regular solution of Eq. (\ref{eqf}) is $Z_m(x)= P_m (1/x)$. Therefore, the
general solution of the problem regular at the Alfven surface
can be presented  as follows
\begin{equation}
f=\sum_m a_m P_m (1/x)Q_m (\eta).
\label{sol}
\end{equation}
The boundary conditions on the stellar surface define the numerical
coefficients $a_m$.

It is convenient to take one term from expansion (\ref{sol}) which
corresponds to the simplest non uniformity of the magnetic field on
the stellar surface. It is especially interesting to consider the
non uniformity which
increases the magnetic field at the magnetic poles and
decreases the magnetic field at the magnetic equator. The term with $m=2$
 corresponds to this distribution. The correction to the solution in this
case is
\begin{equation}
f= {(3/x^2-1)\over (3/x^2_*-1)}\sin^2\chi\cos\chi.
\label{eq47}
\end{equation}
The corrections to the magnetic field are
\begin{equation}
h_{\chi,1}= {6H_*R_*^2\over R_a^2 x^4 (3/x_*^2-1)}\sin\chi\cos\chi,
\label{eq48}
\end{equation}
\begin{equation}
h_{r,1}= {(3/x^2-1)H_*R_*^2\over r^2 (3/x_*^2-1)}(2\cos^2\chi-\sin^2\chi).
\label{hr}
\end{equation}
It is seen that the radial perturbation of the magnetic field is positive
at the magnetic poles and negative at the equator (at $\varepsilon > 0$). 
It is interesting
that the perturbation of the radial component of the magnetic field decreases
  as $r^{-2}$ at  $r\rightarrow\infty$ . The initial magnetic field
has the same dependence. This means that  the flow is
not spherically symmetric at $r\rightarrow\infty$.
This is one of the features of supersonic
flows.  The magnetic forces decrease with distance so fast that the
supersonic plasma
moves ballistically along straight lines defined by the initial conditions.
The magnetic pressure gradient across the flow lines is not able to 
change the trajectories of the particles of the plasma to restore the
spherical symmetry at $r\rightarrow\infty$.
\begin{figure}
\centerline{\psfig{file=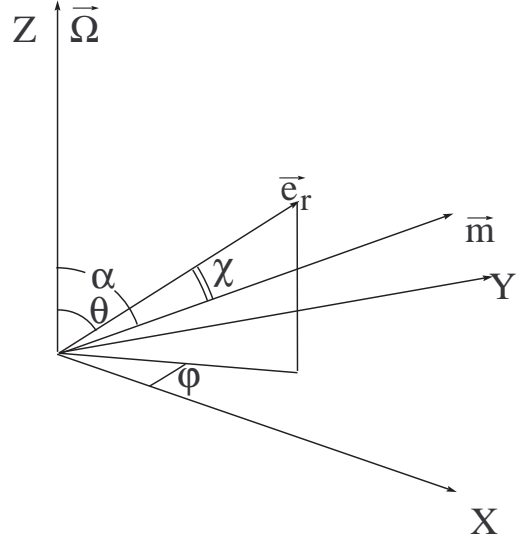,width=7.0truecm,angle=0}}
\caption{The geometry of the coordinate systems connected with the axis of
rotation (axis ${\bf \Omega}$) and the axis of symmetry of the magnetic field
(axis ${\bf m}$) inclined to the rotation axis ${\bf \Omega}$ under the angle
$\alpha$. ${\bf e}_r $ is the unit vector directed along ${\bf r}$.}
\label{fig1}
\end{figure}

The components of the arbitrary  vector $\bf A$
in the  spherical coordinate system with the polar axis
directed along the symmetry axis of the
magnetic field ${\bf m}$ (dashed coordinate system) can be transformed
into the components of the same vector in
 the laboratory coordinate system,   shown in
Fig. \ref{fig1} as follows:
$$
A_r=A_r^{'} ,
$$

$$
A_{\theta}={A_{\chi}^{'} (\cos\alpha\sin\theta-\sin\alpha\cos\theta\cos\varphi)
-A_{\varphi}^{'} \sin\varphi\sin\alpha\over\sin\chi},
$$

$$
A_{\varphi}={A_{\varphi}^{'} (\cos\alpha\sin\theta-
\sin\alpha\cos\theta\cos\varphi)+
A_{\chi}^{'} \sin\varphi\sin\alpha\over\sin\chi},
$$
where $\alpha$ is the inclination angle of the magnetic axis in relation to
the axis $\bf\Omega$. The equation connecting the coordinates in the
 systems is as follows
\begin{equation}
\cos\chi=\sin\alpha\sin\theta\cos\varphi+\cos\alpha\cos\theta.
\label{coschi}
\end{equation}

The 
transformation of the vector ${\bf A}$ with
$A_{\varphi}=0$ at $\alpha=\pi/2$ (orthogonal rotator)
is of special interest for us below.
The  components of the
vector are connected as  follows in this case
\begin{equation}
A_{\theta}=-A_{\chi}^{'}{\cos\theta\cos\varphi\over
\sqrt{1-\sin^2\theta\cos^2\varphi}}
\label{tr1}
\end{equation}
and
\begin{equation}
A_{\varphi}=A_{\chi}^{'}{\sin\varphi\over\sqrt{1-\sin^2\theta\cos^2\varphi}}.
\label{tr2}
\end{equation}

\subsection{Plasma acceleration in the slightly nonaxisymmetric magnetic 
field at slow rotation}

At slow rotation and  small nonaxisymmetry of the magnetic field the
solution can be expanded in two  parameters: $\Omega$ and
$\varepsilon$. The expansion takes a form
\begin{equation}
{\bf H}_p={\bf H}_{p,0}+\varepsilon{\bf h}_{p,1}+ ....,
\end{equation}
\begin{equation}
H_{\varphi}=\Omega H_{\varphi,1}+\varepsilon h_{\varphi,1}+ ....,
\label{tor}
\end{equation}
\begin{equation}
{\bf v}_p ={\bf v}_{p,0}+\varepsilon{\bf V}_{p,2}+ ....,
\end{equation}
\begin{equation}
v_{\varphi}=\varepsilon V_{\varphi,3}+ ....,
\end{equation}
\begin{equation}
n= n_{0}+\varepsilon n_1 + ....,
\end{equation}
\begin{equation}
\gamma=\gamma_0+\varepsilon\Omega \delta\gamma  + .......
\end{equation}
It follows from (\ref{gamma}) that
$\delta\gamma$ satisfies the equation
\begin{equation}
mn_0c^2\Omega\varepsilon v_0{\partial\delta\gamma\over \partial r}=
E_\theta j_{\theta},
\label{57}
\end{equation}
where
\begin{equation}
j_{\theta}=\varepsilon {c\over 4\pi}
({1\over r\sin\theta}{\partial h_{r,1}\over \partial\varphi}-
{\partial\over r\partial r}r h_{\varphi,1}),
\label{eqg}
\end{equation}
and $E_{\theta}$ is defined by Eq. (\ref{e1}).
Let us calculate $\delta\gamma$ for the particular case of  orthogonal
rotation ($\alpha=\pi/2$) for the flow in the equatorial plane
($\theta=\pi/2$). The transformation of  the solution (\ref{tr1}, \ref{tr2})
to the laboratory coordinate system gives
\begin{equation}
j_{\theta}={c\over 4\pi}{6\varepsilon H_*R_*^2\sin\xi\cos\xi\over
r^3(3/x_*^2-1)}.
\label{59}
\end{equation}
Integration of Eq. (\ref{57})  with Eq. (\ref{59}) gives
\begin{equation}
\gamma=\gamma_0-\gamma_0{\Omega R_*\over c}{v_0\over c}
{6\varepsilon\sin\xi\cos\xi\over
(3-x^2_*)},
\label{lorentz}
\end{equation}
where
\begin{equation}
\xi=\varphi-\Omega t.
\label{eq63}
\end{equation}
It is seen  that the violation of axial symmetry of the magnetic
field  essentially modifies  the acceleration process. 
The first correction to the Lorentz factor
of the plasma  is proportional to $\Omega^4$ in the axisymmetric
magnetic field,
while even the small nonaxisymmetry of the magnetic field makes the
dependence  much  stronger. The first correction
to the Lorentz factor begins with the term  proportional to $\Omega$.

\begin{figure}
\centerline{\psfig{file=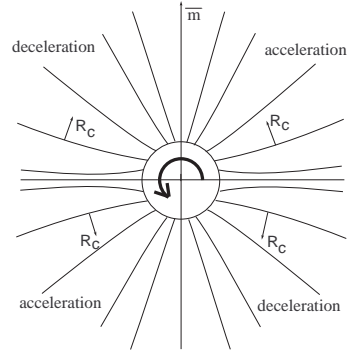,width=5.0truecm,angle=0}}
\caption{Outflow in  the magnetic field of
a star with a monopole-like magnetic field having  small 
non uniformity on the surface of the star. The magnetic field is slightly 
stronger at the poles and slightly weaker at the magnetic equator.
The magnetic field is axially symmetric in relation to the axis $\bf m$. 
The particular case when the axis of the symmetry of the
magnetic field is  perpendicular to the axis of rotation (orthogonal
rotator) is presented.  At a slow rotation, the acceleration
of  plasma occurs along  the field lines which curve with a radius $R_c$ 
directed into the direction of rotation shown by thick arrow.}
\label{fig2}
\end{figure}
The schematic structure of the field lines unperturbed by the rotation
is shown in Fig. \ref{fig2}. The view from the top is presented. The
acceleration of the plasma takes place in the sectors where
$\sin\varphi\cos\varphi < 0$ (at $t=0$, $\varphi=0$ at the direction
along ${\bf m}$).
In other sectors the plasma is decelerated.
It follows from the figure that
the plasma is accelerated where the field lines unperturbed
by the rotation are curved into  the direction of the rotation and
is decelerated where they are curved in the opposite
direction. This behaviour of the plasma has a simple mechanical analogy.

The motion of the magnetised plasma can be considered in some sense as a
motion of a bead on a wire (Blandford \& Payne \cite{bp}).
The bead is accelerated centrifugally during  the motion along the wire,
 provided
that the wire is not in the  shape of an
 Archimedian spiral with the components
$dl_{\varphi}$ and $dl_r$ defined by the equation
$$
{dl_{\varphi}\over dl_r}=-{r\Omega\sin\theta\over v_r}.
$$
It is easy to demonstrate that any departure of the wire shape from
the Archimedian spiral results in variation of the energy of the bead.
The bead will be accelerated centrifugally under
the condition $|{dl_{\varphi}\over dl_r}| < |{r\Omega \sin\theta\over v_r}|$
and decelerated in the opposite case.
The plasma behaves exactly in accordance with this mechanical model.
It follows from (\ref{bogsol}) that the  field lines of the axisymmetric
magnetic field 
form the Archimedian spiral at small $\Omega$.
At the nonaxisymmetric outflow, the toroidal magnetic field is a linear
superposition of perturbations from rotation and the nonaxisymmetry. 
The initial nonaxisymmetry  straightens the magnetic field lines in the
sectors where
$\sin\varphi\cos\varphi < 0$. The shape of the field lines here satisfies
the  inequality
$|{dl_{\varphi}\over dl_r}| < |{r\Omega \sin\theta\over v_r}|$. The plasma
is accelerated in this case.
In other sectors, the field lines are curved in an azimuthal direction
more strongly in comparison to  the
Archimedian spiral. Therefore, the plasma is decelerated.

\section{Generation of magneto sound waves}

The modification of the MC acceleration is not the only
new  feature of the outflows in the
magnetic field with  violated axial symmetry.  
The source with the
nonaxisymmetric magnetic field excites the
magnetohydrodynamical (but not magnetodipole) waves in the
wind. Perturbation theory allows us to investigate
this process while the amplitude of these waves is small.

In the axisymmyteric case,
the azimuthal magnetic field generated by the rotation is defined by 
Eq.(\ref{bogsol}).
Perturbation theory can  be applied
while the pressure of the toroidal
magnetic field in the co-moving coordinate system is less than the
pressure of the unperturbed poloidal magnetic field
(Bogovalov \cite{bog97}). In the laboratory coordinate system
this condition takes the form
\begin{equation}
H_{\varphi}^2-E^2 < H_p^2,
\end{equation}
This inequality limits the application of the perturbation theory by the
distances $r$ less than
$\gamma_0 v_0/\Omega$.  When the last value  becomes less than $R_a$,
the perturbation becomes strong everywhere in the flow.
Thus, the perturbation theory can be applied only if
\begin{equation}
{\Omega R_a\over \gamma_0 v_0} < 1.
\label{cond1}
\end{equation}
The objects satisfying   condition (\ref{cond1}) are referred to as  slow
rotators. In the opposite case they are  fast rotators.
Analysis shows that  all radio pulsars, including the Crab pulsar, are slow
rotators at the conventional parameters of the winds (Bogovalov \cite{bog99}).

Eq. (\ref{cond1}) defines the condition under which the perturbation of the flow 
in the subsonic region remains small compared with the initial magnetic field. 
Our use of the perturbation theory is valid given an additional 
restriction. During the calculation of the first corrections, we neglected the 
contribution of the energy density of the toroidal and electric fields
generated by the rotation into the
inertial mass of the plasma. In order words, it was assumed that the 
condition
\begin{equation}
mc^2n^{'} \gg {H^{'2}_{\varphi}\over 4\pi}
\end{equation}
is fulfilled, 
where the upper index ``$^{'}$'' denotes values in the co-moving coordinate 
frame.
It results in the additional condition  
\begin{equation}
R_a \ll R_L,
\label{cond2}
\end{equation}
where $R_L=\Omega/c$ is the radius of the light cylinder. 
Thus, our consideration is valid under the conditions (\ref{cond1}) and 
(\ref{cond2}) in the region limited by the radius $\gamma_0 v_0/ \Omega$.

The objects with  weak violation of the axial symmetry excite small amplitude 
MHD waves in the outflowing wind, which can be considered in the framework of 
the 
perturbation theory. These waves are 
formed in  the wave zone with dimensions of the
wavelength $\lambda=V_w/\Omega $  (Landau \& Lifshitz \cite{landauhydro}),
where $V_w$ is the velocity of the wave  propagating in the moving plasma.
The fast magnetosonic velocity of the cold wind decreases with
an increase in $r$.
In the region $r > R_a$, the fast magnetosonic velocity of the plasma $v_f$ is  
small
compared with $v_0$ for slow rotators and the condition $v_0 \gg v_f$ is 
fulfilled everywhere along the flow, under  condition (\ref{cond2}). 
The wavelength can be  
estimated as $\lambda \approx v_0/\Omega$ in this region. Thus,
four scales appear in the problem.
One is the initial radius of the Alfven (or fast magnetosonic) surface, $R_a$.
Others are the  radius of the wave zone $\lambda$, the radius of 
the light cylinder $R_L$ and  the radius
$r_{max} =\gamma_0\lambda$, where the perturbation theory itself can not be
applied. 
The inequality  $R_a/\gamma_0\lambda < 1$ is fulfilled for the slow rotators.
For the nonrelativistic plasma, $r_{max}=\lambda$. The wave zone is located in 
the region where  perturbation theory 
can not be applied. Therefore, the generation
of MHD waves in the nonrelativistic winds cannot be considered 
in the framework of the perturbation theory developed here. 
However, in the relativistic plasma with $\gamma_0 \gg 1$
there is a gap between
$\lambda$ and $\gamma_0 \lambda$ where the perturbation theory remains valid.
The MHD wave generation can be considered in the framework of this theory.

It is easy to extend our  solution to the zone
$ \lambda < r < \gamma_0\lambda$,  following  Landau \& Lifshitz
(\cite{landauhydro}, sect. 74).
The electric and magnetic fields in the outgoing wave depend on $r$
and $t$ as $A_i(t-r/v_0)/r$, where $A_i$ are  unknown functions. 
In the  zone $R_* \ll r < \lambda$ the solution  has a form $B_i(t)/r$. Both 
solutions must coincide  at $r = \lambda$. This implies that
$A_i = B_i(t-r/v_0)$. So,  we need only to define functions $B_i$
of the solution obtained in the zone $r < \lambda$ to obtain the solution at
$r > \lambda$.

It was pointed out that the perturbation of the radial component of the  
magnetic field $h_r$, proportional to $\varepsilon$,  falls down as $r^{-2}$ at 
large $r$.
This perturbation  results in the perturbation of the toroidal magnetic field
(see Eq. (29))
\begin{equation}
h_{\varphi}= -{\Omega r\sin\theta\over v_0} h_r
\label{hfp}
\end{equation}
and in the perturbation of the $\theta$ component of the electric 
field (see Eq. (20))
\begin{equation}
e_{\theta}=-{\Omega r\sin\theta\over c} h_r.
\label{ep}
\end{equation}
The electric and magnetic fields from (\ref{hfp}) and  (\ref{ep})  decrease
with distance as $r^{-1}$ at $r \gg R_*$, since $h_r$ has the following
dependence ( see Eq. (49))
\begin{equation}
h_r=-\varepsilon {H_*R^2_*\over r^2(3/x_*^2-1)}(3\cos^2\chi-1).
\label{hr2}
\end{equation}
All other terms in Eq. (\ref{hr}) are neglected since they decrease with
$r$ faster than $r^{-2}$.
According to Eq. (\ref{coschi}) and Eq. (\ref{eq63}),
$\cos\chi$ depends on $t$ and $\varphi$
in the combination $\xi =\varphi-\Omega t$.
To obtain the solution  at $r > \lambda $, we have to replace it by 
the combination $\varphi-\Omega (t-t/v_0)$. As a result,
$\cos\chi$ takes the form (see Eq. (\ref{coschi}))
\begin{equation}
\cos\chi=\sin\alpha \sin\theta\cos(\varphi-\Omega(t-
r/v_0))+\cos\alpha\cos\theta.
\label{cos}
\end{equation}

The solution (\ref{hfp},\ref{ep},\ref{hr2},\ref{cos})
allows us to define a correction to the rotational losses of the
central object due to the emission of the MHD waves in the wind and to
estimate the effect of the wind acceleration by these waves.
The correction to the $r$ component of the
Poynting flux ${\bf S}={c\over 4\pi} E\times H$ due to the perturbation of the
electric and magnetic fields is as follows
\begin{equation}
S_r = -\varepsilon v_0{H_*^{2}R_*^2\over 2\pi r^2 (3/x_*^2-1)}
({\Omega R_*\sin\theta\over v_0})^2
(3\cos^2\chi -1).
\end{equation}
The ratio of this correction over the particle flux density in $mc^2$ units
is as follows
\begin{equation}
\delta\sigma=-2\varepsilon ({\Omega R_a\sin\theta\over c})^2\gamma_0
{(3\cos^2\chi -1)\over (3/x_*^2-1)}.
\label{sigma}
\end{equation}
This value shows the maximal possible variation of the Lorentz factor
of the plasma at the full absorption of the MHD wave.

Integration of the correction to the Poynting flux over a surface 
surrounding the  source gives the first correction to the rotational 
losses of the central object,
\begin{equation}
\delta \dot E_{rot}={4\over 15}\varepsilon \Omega^2
{(H_* R_*^2)^2\over v_0 (3/x_*^2-1)}(3\cos^2\alpha-1).
\label{edot}
\end{equation}
The zero order expression for the rotational losses is given in
(Bogovalov \cite{bog99}). As  was expected, the departure from  uniform
distribution of the magnetic field lines on the surface of the star
changes the rotational losses.
Correction (\ref{edot})  depends on the sign of $\varepsilon$ and
does not go to zero even at  axisymmetric rotation ($\alpha = 0$).
This fact has a simple explanation. The total Poynting flux
from an axisymmetrically rotating object
depends on the distribution of the poloidal magnetic field on latitude at
a large distance from the source (Bogovalov \cite{bog99}).
At the fixed total magnetic flux, the rotational losses increase when the
magnetic flux is more concentrated at the equator. The concentration of the
magnetic flux at the poles consequently
results in the  decrease of  rotational loss.
Eq. (\ref{edot}) describes this dependence. The magnetic flux is concentrated
at the poles for $\varepsilon > 0$ on the surface of the star,
but according to (49), at large distances
the magnetic flux is more concentrated  at the
equator. Therefore, the rotational losses increase for $\varepsilon > 0$
and decrease in the opposite case. 

Qualitatively, this also explains the fact that according to Eq. (\ref{edot})
the rotational losses decrease
at the orthogonal rotation (at $\varepsilon > 0$). In this case, the magnetic
flux at large distances
is more concentrated along the plane of the magnetic equator, oriented
perpendicular to the rotational equator. The average
magnetic field in the equatorial plane decreases and  results in the
decrease in rotational losses.

The MHD wave propagating in the wind
 changes the energy of the particles of the plasma. The
variation in the energy is described by Eq. (31) with the electric
current ${\bf j} ={c\over 4\pi}(curl {\bf H} -{1\over c}{\partial
{\bf E}\over \partial t})$. Substitution of  solutions (\ref{hfp}, \ref{ep})
in this expression gives
\begin{equation}
j_{\theta}= {c\over 4\pi}{\Omega^2 r\sin\theta\over v_0^2\gamma_0^2}
{\partial h_r\over\partial\xi}.
\label{jt}
\end{equation}
The correction to the Lorentz factor of the plasma depends on its
arguments, as  $\delta \gamma(r, \xi)$.
After substitution of this expression and (\ref{jt})
in Eq. (31), we obtain
\begin{equation}
{\partial \delta\gamma\over\partial r}=-6\varepsilon R_a^2
{\Omega^3\sin^3\theta\over \gamma_0 v_0 c^2}
{\cos\chi\sin\alpha\sin\xi\over(3/x_*-1)}.
\label{dg}
\end{equation}

Formally, integration of Eq. (\ref{dg}) over $r$ gives an
infinite result. It is
clear why this should occur.
We do not take into account the decrease in the amplitude   of the
MHD wave due to the redistribution of  energy between  waves and particles 
of the plasma. However, we can estimate the variation in the energy of the 
plasma  depending on $r$. Integration of Eq. (\ref{dg}) gives
\begin{equation}
\delta\gamma = -6\varepsilon ({R_a \Omega \sin\theta\over c})^2 
{\sin\theta\cos\chi\sin\alpha\sin\xi\over (3/x_*^2-1)}
{r\Omega\over \gamma_0 v_0}.
\label{int}
\end{equation}
The lower limit of integration is neglected here.
This correction can be  negative or positive, depending on the sign of 
$\cos\chi\sin\xi$. Thus, we consider the process of redistribution of 
energy, but not  pure acceleration.
Comparison of Eq. (\ref{int}) with Eq. (\ref{sigma}) shows that even at the 
distance $\gamma_0\lambda$ only a small part of the energy of the wave is 
transformed into the energy of the plasma and vice versa at
$\gamma_0 \gg 1$.   So, this process is extremely noneffective in the
relativistic regime.

\section{Implications for the physics of
relativistic winds from radio pulsars}

It is generally believed that  $e^{\pm}$ plasma is produced  
and initially accelerated in  the pulsar magnetosphere in the so-called 
electrostatic gaps (Ruderman \& Sutherland \cite{ruderman}, 
Arons \cite{arons}, Cheng et al. \cite{cheng}, Romani \cite{romani}). 
The ideal MHD does  not apply to these gaps. However,  these gaps occupy
only a small
part of the magnetosphere. They produce the initially relativistic plasma,
which is dense enough to screen the electric field and to provide the
ideal MHD
conditions outside the gaps. Therefore, the flow of this plasma  can be 
described in an ideal MHD approximation as  the wind. Below, we
assume for simplicity that the wind  with a prescribed 
initial density and Lorentz factor is formed somewhere near the surface 
of the pulsar.  
  
\begin{figure}
\centerline{\psfig{file=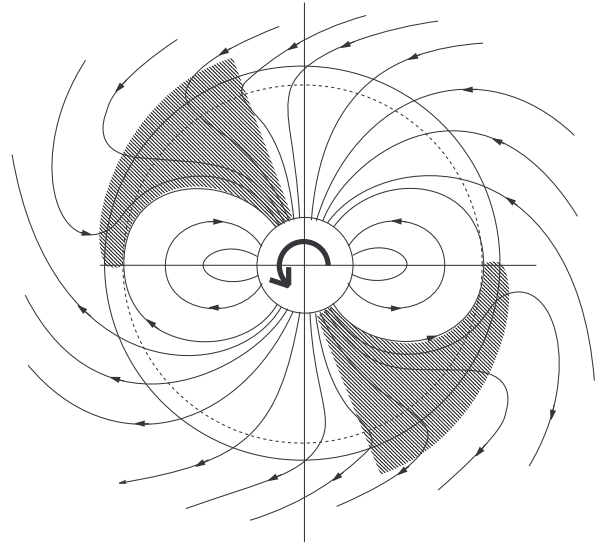,width=8.0truecm,angle=0}}
\caption{The view from the top of the schematic magnetic field structure of
the outflow from an orthogonal rotator  with an initially dipole
magnetic field. The light cylinder is shown by the solid circle.
Direction of rotation is shown by the thick arrow.
The region of the closed field lines does not reach
the light cylinder and is
limited by the dashed circle. Some  of the open field 
lines (shadowed) are curved
in the direction of rotation  up to the light cylinder.
The magnetocentrifugal acceleration will take place on these field lines.
Beyond the light cylinder, all the open field lines are curved in the
Archimedian spiral typical of  axisymmetric outflows.}  
\label{fig3}
\end{figure}

The schematic structure of  an
orthogonal rotator with an initially dipole magnetic field is presented
in Fig. \ref{fig3}.
The dipole magnetic field is not
axisymmetric at   oblique rotations.
There are regions of closed field lines in the
magnetosphere and regions of open field lines along which the plasma
outflows. Some of these
field lines are initially  curved in the   direction of rotation.
Therefore, the new type of 
MC plasma acceleration  discussed in this paper
 must work under radio pulsar conditions.
The question remains, however, whether 
this mechanism of plasma acceleration essentially
changes the energy of the plasma in pulsar conditions.

Real pulsar parameters corresponds to the conditions $R_a > R_L$ and
$R_a < \gamma_0 R_L$. So, our solution cannot be directly applied to radio 
pulsars since it is valid at $R_a < R_L$. However, this solution
can give important qualitative information about the process of 
wind acceleration in any magnetosphere having violated the condition of
axial symmetry.

To understand at least qualitatively  the influence of the violation
of the axial symmetry  on the process of the
MC acceleration  in a pulsar magnetosphere we should estimate
the dimensionless values of the expansion parameters in Eq.
(\ref{lorentz}) for pulsar conditions. If these parameters are
small, it would imply that this effect is not important for  radio
pulsars. If these parameters are of the order of 1, it would imply that 
the MC acceleration  in the pulsar conditions can
essentially change the energy of the plasma.

The expansion in Eq. (\ref{lorentz}) occurs in two parameters $\varepsilon$ and
${\Omega R_*\over c}$. $\varepsilon$ can be estimated as 
$\varepsilon \approx \delta H_*/<H_*>$ where $\delta H_*$ is the maximal 
variation of the magnetic field on the surface of the star and $<H_*>$ is the 
average magnetic field on the surface of the star. It is evident that 
$\varepsilon \sim 1$ for dipolar magnetic fields which are believed 
to occur on the  pulsar surface.

In the second dimensionless parameter ${\Omega R_*\over c}$,
the physical sense
of  $R_*$ is the most unclear.  At first glance, it seems that this 
parameter defines the scale over which the magnetic field decreases with 
distance from the star. However, more
close inspection of Eq. (\ref{lorentz}) shows that this cannot be correct. 
The first correction to the Lorentz factor (\ref{lorentz})
does not depend on the
magnetic field $H_*$ at all, provided that   $3 R_a^2 \ll R_*^2$.
Therefore,  $R_*$ cannot be interpreted as describing
the scale of decrease of the magnetic field.

To understand the physical sense of $R_*$ in  Eq. (\ref{lorentz}),
it is necessary to note that the  
MC  acceleration under  consideration occurs only 
in  regions where the magnetic field lines were initially curved into
the direction
of rotation. Simple analysis of the correction to the magnetic field 
(\ref{eq47}) 
shows that the violation of axial symmetry on the surface of the star 
results in   the bending of  field lines in the region with a
characteristic size
$R_*$. The shape of the field lines does not depends on $H_*$ at
$3 R_a^2 \ll R_*^2$. This inequality means that the energy density of the
magnetic field exceeds the energy density of the plasma near the star
surface.
 At larger distances,
the field lines straighten
and do not accelerate the plasma.  Thus, $R_*$ 
plays the  role of the characteristic scale $R_{curve}$ on which  the magnetic 
field lines have nonzero curvature.  Only accidentally does
$R_{curve} = R_*$
in our specific case. Thus, the second parameter defining the expansion 
of the terminating Lorentz factor can be presented as $\Omega R_{curve}/c$.

In the pulsar magnetosphere, the scale of the poloidal 
magnetic field variation is equal to the radius of the star, $R_*$.
The scale
$R_{curve}$ depends on the topology of the poloidal magnetic field. In the 
conventional models, the
region of the closed field lines almost reaches the light cylinder. Direct
calculations in the axisymmetric model
confirm this assumption (Contoupolos et al. \cite{janis}).
For this magnetic field topology
$R_{curve} \sim R_L$ and therefore, the  parameter $\Omega R_{curve}/c$
also should be close to 1. 
For this reason, we conclude that the influence of the 
violation of the magnetic field axial symmetry on the process of 
the  MC acceleration  is certainly not small in the pulsar
conditions and it could provide efficient acceleration of the 
relativistic winds.

It is difficult to estimate reliably the terminating Lorentz
factor of the plasma accelerated due to this mechanism in the
magnetosphere of pulsars. Any field line initially curved into
the direction of rotation is transformed
into an Archimedian spiral well beyond the light cylinder where 
acceleration does not occur. The maximal Lorentz factor is defined 
by the distance where this transition occurs. This is defined 
by the  field structure at the vicinity of the light
cylinder near the last closed field line of the magnetosphere, as  is
shown in Fig. \ref{fig3} (shadowed region).
Only accurate mathematical solutions will give an accurate answer
to  questions regarding the terminating Lorentz factor of the plasma.
However, it is easy to understand that if this mechanism really provides
transformation of the  Poynting flux into kinetic
energy of the plasma, then the acceleration must take place very close to the
light cylinder.

In conclusion, we note that the acceleration of the
relativistic
winds  at the light cylinder of radio pulsars looks attractive from 
an astrophysical point of view as well.
The mechanism of MC acceleration of the plasma which we have described 
can allow us to resolve a long-standing problem of  additional
acceleration of the wind from the Crab pulsar. It follows
from the comparison of the observations of the Crab Nebula
with the MHD theory of the
interaction of the pulsar wind with the interstellar medium that the
ratio of the Poynting flux to the density of the kinetic energy flux in
the wind before the terminating shock 
 is of the order $\sim 10^{-3}$ 
(Rees \& Gunn \cite{rees}, Kennel \& Coroniti \cite{kennel}). 
However,  this ratio cannot be achieved
at the dissipativeless axisymmetric outflow(see in particular Chiueh et al.'s 
(1998)
discussion of this question). It was argued recently by Begelman (1998) that
the ratio of the Poynting flux to the kinetic energy flux of the order
of 1 can also be consistent with the observations of the Crab Nebula if we
take into account a possible instability of the flow after the terminating
shock. In any case, an additional 
acceleration of the wind of the 
relativistic plasma is needed  to explain the  observations of the 
Crab Nebula, since the electrostatic gap models are not able to reproduce a
wind with the necessary characteristics (Arons \cite{arons96}). If we assume  
that  the wind is accelerated due to the MC mechanism 
and that it carries off
approximately half of the total rotational energy of the pulsar
in the form of the kinetic energy, this
resolves the problem of  wind acceleration. The
ratio of the Poynting flux to the kinetic energy flux of the plasma appears
to be of the order of 1 immediately after the light cylinder, which is
consistent with the observations of the
Crab Nebula according to  Begelman's scenario (1998). This also resolves 
the difficulties with the mechanism of the wind acceleration
at large distances
from the pulsar proposed by  Coroniti (1990) and recently revisited by 
Lyubarskii \& Kirk (\cite{kirk00}). There is simply no need to invoke
additional mechanisms in this acceleration.

Analysis performed by Aharonian \& Bogovalov
(1999) (see also Bogovalov \& Aharonian (2000)) shows that the wind from the 
Crab pulsar with 
energetics comparable to the total rotational losses of this pulsar
and accelerated near the light cylinder
should have a total particle flux of $\sim 10^{40}$part./s and an
average Lorentz factor of $\sim 10^5$ to be consistent with the limitations
imposed by observations of the Crab Nebula in VHE gamma-rays.
It is interesting that the wind with
these parameters indeed provides the total injection rate into the Crab
Nebula necessary to explain  the electromagnetic emission 
in the wide range from radio to  X-rays (Rees \& Gunn \cite{rees}). Thus, the 
idea
that the mechanism of  magnetocentrifugal acceleration of the plasma 
in the axially non uniform magnetic field can be responsible for the 
acceleration of the wind in the magnetosphere of radio pulsars does not 
contradict  observations of the Crab pulsar and Nebula and deserves 
further development in a more realistic model.

\begin{acknowledgements}
This work was supported partially by the Russian Ministry of Education in
the framework of the program ``Universities of Russia - basic research",
project N 990479 and by INTAS-ESA grant N99-120. The author is grateful
for fruitful discussions with the
participants of the seminar of the relativistic astrophysics group
at the  Sternberg Astronomical Institute of the
Moscow  State University.
\end{acknowledgements}

\end{document}